\documentclass[conference]{IEEEtran}
\IEEEoverridecommandlockouts

\usepackage{cite}
\usepackage{amsmath,amssymb,amsfonts}
\usepackage{algorithmic}
\usepackage{graphicx}
\usepackage{textcomp}
\usepackage{xcolor}
\def\BibTeX{{\rm B\kern-.05em{\sc i\kern-.025em b}\kern-.08em
    T\kern-.1667em\lower.7ex\hbox{E}\kern-.125emX}}
\begin{document}

\title{Pushing the Limit of Sound Event Detection with \\ Multi-Dilated Frequency Dynamic Convolution
\thanks{This work was supported by the Institute of Civil Military Technology Cooperation funded by the Defense Acquisition Program Administration and Ministry of Trade, Industry and Energy of Korean government under grant No. UM22409RD4, and Korea Research Institute of Ships and Ocean engineering a grant from Endowment Project of “Development of Open Platform Technologies for Smart Maritime Safety and Industries” funded by Ministry of Oceans and Fisheries(PES4880).}
}

\author{\IEEEauthorblockN{Hyeonuk Nam}
\IEEEauthorblockA{\textit{Department of Mechanical Engineering} \\
\textit{Korea Advanced Institute of Science and Technology}\\
Daejeon, South Korea \\
frednam@kaist.ac.kr}
\and
\IEEEauthorblockN{Yong-Hwa Park}
\IEEEauthorblockA{\textit{Department of Mechanical Engineering} \\
\textit{Korea Advanced Institute of Science and Technology}\\
Daejeon, South Korea \\
yhpark@kaist.ac.kr}
}

\maketitle

\begin{abstract}
Frequency dynamic convolution (FDY conv) has been a milestone in the sound event detection (SED) field, but it involves a substantial increase in model size due to multiple basis kernels. In this work, we propose \textit{partial frequency dynamic convolution (PFD conv)}, which concatenates outputs by conventional 2D convolution and FDY conv as static and dynamic branches respectively. PFD-CRNN with proportion of dynamic branch output as one eighth reduces 51.9\% of parameters from FDY-CRNN while retaining the performance. Additionally, we propose \textit{multi-dilated frequency dynamic convolution (MDFD conv)}, which integrates multiple dilated frequency dynamic convolution (DFD conv) branches with different dilation size sets and a static branch within a single convolution layer. Resulting best MDFD-CRNN with five non-dilated FDY Conv branches, three differently dilated DFD Conv branches and a static branch achieved 3.17\% improvement in polyphonic sound detection score (PSDS) over FDY conv without class-wise median filter. Application of sound event bounding box as post processing on best MDFD-CRNN achieved true PSDS1 of 0.485, which is the state-of-the-art score in DESED dataset without external dataset or pretrained model. From the results of extensive ablation studies, we discovered that not only multiple dynamic branches but also specific proportion of static branch helps SED. In addition, non-dilated dynamic branches are necessary in addition to dilated dynamic branches in order to obtain optimal SED performance. The results and discussions on ablation studies further enhance understanding and usability of FDY conv variants.
\end{abstract}

\begin{IEEEkeywords}
sound event detection, frequency dynamic convolution, partial frequency dynamic convolution, multi-dilated frequency dynamic convolution
\end{IEEEkeywords}

\section{Introduction}
\label{sec:intro}
The advent of deep learning has spurred extensive research in sound event detection (SED) \cite{CASSE, DCASEtask4}. Initially, SED development borrowed methods from other domains but has since evolved with techniques tailored specifically for SED \cite{dcase2020_1st, dcase2021_1st, dcase2023a_1st, FDY, mdfdy, astsed, freqatt}. Among these, frequency dynamic convolution (FDY conv) represents a significant advancement in the field, as it was adopted to many high-ranked submissions to detection and classification of acoustic scenes and events (DCASE) challenge SED task \cite{dcase2022_2nd, dcase2023a_1st, dcase2023a_2nd, dcase2024mytechrep}. By reducing the translational equivariance of 2D convolution in the frequency dimension and adapting convolution kernels to input content, FDY conv has greatly enhanced SED performance and inspired follow-up studies \cite{dcase2023a_1st, mdfdy, dcase2023a_2nd, STRF, DFD}.

However, FDY conv involves a substantial increase in model size due to multiple basis kernels \cite{FDY, dyconv}. FDY conv with four basis kernel almost triples the parameter count in convolutional recurrent neural network (CRNN) architecture. Thus lighter alternative is needed for more efficient SED. In this context, time-frame frequency-wise squeeze and excitation has been proposed \cite{freqatt}, but it does not solve translational equivariance problem of 2D convolution on frequency dimension.

To minimize the parameter increase in FDY conv, we propose \textit{partial frequency dynamic convolution (PFD conv)}. By  concatenating conventional 2D convolution output and FDY conv output, PFD conv effectively reduces the model size while retaining the performance. Furthermore, by introducing multiple dynamic branches, we can employ multiple independent FDY convs in single convolution layer. To further enhance the performance, multiple dilated frequency dynamic convolution (DFD conv) modules with different dilation size sets are added to the static convolution branch \cite{DFD}. Resultant \textit{multi-dilated frequency dynamic convolution (MDFD conv)} achieves state-of-the-art performances on domestic environment sound event detection (DESED) dataset with and without external dataset. The main contributions of this paper are as follows:
\begin{enumerate}
    \itemsep 0.1em
    \item{We introduced a concept of concatenating static and dynamic convolution branches for efficient and powerful SED.}
    \item{Extensive ablation studies showed various proportions of static and dynamic branches with varying dilation sizes to balance model size and performance.}
    \item{Proposed \textit{partial frequency dynamic convolution (PFD conv)} reduces the model size by 51.9\% while retaining the performance of FDY conv.}
    \item{Proposed \textit{multi-dilated frequency dynamic convolution (MDFD conv)} aggregates various DFD conv branches to outperform FDY conv by 3.17\%.}
\end{enumerate}
The official implementation code is available on GitHub\footnote{https://github.com/frednam93/MDFD-SED}.

\section{Methods}
\label{sec:methods}
\begin{figure}[t]
\centerline{\includegraphics[width=8.5cm]{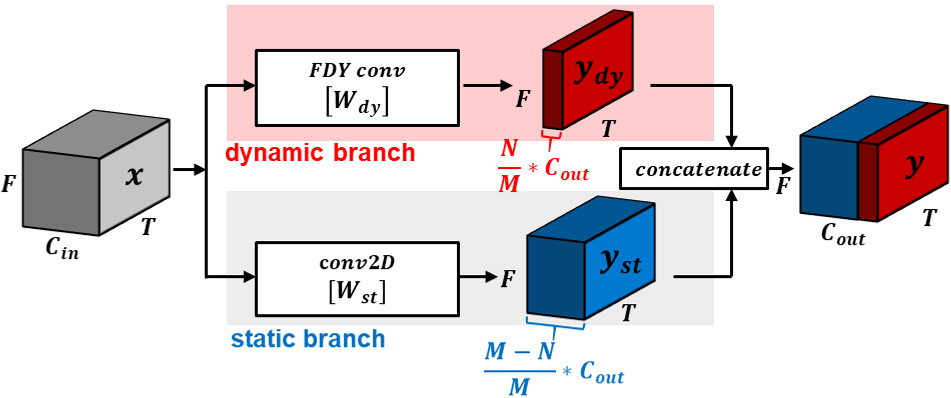}}

\caption{An illustration of partial frequency dynamic convolution (PFD conv) operation which consists of dynamic FDY conv branch and static conventional 2D convolution branch. $x$ and $y$ are input and output of PFD conv. $N$ and $M$ determine the proportion of output of which FDY conv is applied.}
\label{fig:PFDconv}
\end{figure}

\subsection{Partial Frequency Dynamic Convolution}
Partial frequency dynamic convolution (PFD conv) concatenates outputs by FDY conv (dynamic branch) and conventional 2D convolution (static branch) along channel dimension. Since only a part of PFD conv output channels are obtained from FDY conv, the parameter count is proportionally reduced. While translational equivariance along frequency dimension by 2D convolution was pointed out to be the main cause of limiting SED performance \cite{FDY}, it could somewhat benefit SED considering that neighboring frequency bins share similar time-frequency patterns. This might also be potential reason why using limited number of basis kernel for FDY conv is optimal for SED even though theoretically more basis kernels should improve performance more with more degree of freedom on frequency-adaptive convolution kernels. Thus, proposed PFD conv can maintain or enhance SED performance while reducing the model size not only thanks to the dynamic branch but also static branch. Fig. \ref{fig:PFDconv} illustrate the procedure for PFD conv, where dynamic branch processes FDY conv while static branch processes conventional 2D convolution then the output is concatenated. $T$, $F$, $C_{in}$ and $C_{out}$ describes size of time, frequency, input channel and output channel dimensions.

\subsection{Multi-Dilated Frequency Dynamic Convolution}
\begin{figure}[t]
\centerline{\includegraphics[width=8.5cm]{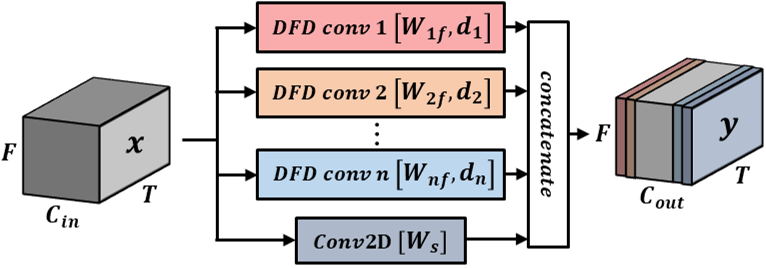}}

\caption{An illustration of multi-dilated frequency dynamic convolution operation. It involves multiple dynamic DFD conv branches and single static branch.}
\label{fig:MDFDconv}
\end{figure}

To further improve SED performance, we could use multiple independent dynamic branches. In addition, we could adopt dilated frequency dynamic convolution (DFD conv) for dynamic branches to diversify and expand frequency-adaptive kernels of dynamic branches \cite{DFD}. Therefore, we propose Multi-dilated frequency dynamic convolution (MDFD conv). As illustrated in Fig. \ref{fig:MDFDconv}, we could apply multiple dynamic branches consisting of DFD conv and single static branch. While multiple static branches could be applied as well, which would be equivalent to group convolution, it is out of the scope for this work. While MDFD conv would involve more parameters than PFD conv does, it could further enhance SED performance with multiple dynamic branches with various dilation size on basis kernels.

\section{Experimental Setups}
\label{sec:setup}
\subsection{Implementation Details}
Domestic environment sound event detection (DESED) dataset is used to train, validate and test SED models in this work. DESED is composed of synthesized strongly labeled dataset, real weakly labeled dataset, real unlabeled dataset and real strongly labeled dataset \cite{DCASEtask4}. No external dataset or external pre-trained model is used in this work except for Table \ref{tab:dcase}, where real strongly labeled dataset is included for comparison with state-of-the-art SED models. We extract mel spectrograms from audio data and input them into CRNN-based SED models. The CRNN models in this work consists of 7 convolution layers, where the first convolution layer uses conventional 2D convolution and the rest convolution layers use FDY conv or its variants. To additionally apply FDY conv on first convolution layer, we introduce pre-convolution before the first convolution for experiments on Table \ref{tab:preMDFD}. At the last convolution layer, DFD conv modules are replaced by non-dilated FDY conv modules as its input only two frequency bins thus dilation along frequency bin do not benefit the pattern recognition \cite{DFD}. Data augmentation methods applied in this work includes frame shift \cite{DCASEtask4}, mixup \cite{mixup}, time masking \cite{specaug} and FilterAugment \cite{filtaug}. We applied median filter with 7 frames (corresponding to 450ms) for all classes as post processing unless stated otherwise.

\subsection{Evaluation Metrics}
To evaluate SED performance, polyphonic sound detection score (PSDS) was used \cite{PSDS}. For DCASE challenge 2021, 2022 and 2023 task4 \cite{DCASEtask4}, two types of PSDS were used. Among them, PSDS2 is more specific to audio tagging than SED, thus we only used PSDS1 in this work \cite{mytechreport, SEBB}. PSDS results listed in the tables are the best score from 12 separate training runs.

\section{Results and Discussion}
\label{sec:result}
\subsection{Partial Frequency Dynamic Convolution}
The results of PFD conv using various proportions is shown in Table \ref{fig:PFDconv}. From the table, PFD-CRNN ($1/N$) denotes that $1/N$ of output channel is obtained by dynamic branch and the rest of the channels is obtained by static branch. We experimented PFD conv with $1/32$, $1/16$ and $n/8$ where $n=1, 2, ..., 7$ as the number of channels on CNN module in CRNN is multiple of 32; 32, 64, 128, 256, 256, 256, 256 from 1st to 7th convolution layers. Note that FDY-CRNN is equivalent to PFD-CRNN with proportion $8/8$. The results show that with a proportion above $1/8$, PFD-CRNN's PSDS is slightly worse or similar to that of FDY-CRNN, except for the model with a proportion $5/8$. But PFD-CRNNs with proportion of $1/16$ and $1/32$ perform worse than FDY-CRNN. The most efficient model is PFD-CRNN with proportion of $1/8$, which only introduce 22.0\% of additional parameters to CRNN and reduces 51.9\% of parameters from FDY-CRNN while retaining the performance by FDY-CRNN.
\begin{table}[t]
\caption{Performance of partial frequency dynamic convolution models with varying proportion of dynamic branch.}
\centering
\begin{tabular}{c|c|c}
\hline
\textbf{models} & \textbf{Params(M)} & \textbf{PSDS1}\\ 
\hline
CRNN            &  4.428             & 0.410          \\ 
PFD-CRNN (1/32) &  4.794             & 0.436          \\
PFD-CRNN (1/16) &  4.996             & 0.434          \\ 
PFD-CRNN (1/8)  &  5.401             & \textbf{0.442} \\
PFD-CRNN (2/8)  &  6.209             & 0.439          \\ 
PFD-CRNN (3/8)  &  7.018             & \textbf{0.443} \\
PFD-CRNN (4/8)  &  7.827             & 0.439          \\
PFD-CRNN (5/8)  &  8.635             & 0.436          \\
PFD-CRNN (6/8)  &  9.444             & 0.441          \\
PFD-CRNN (7/8)  & 10.253             & \textbf{0.443} \\
FDY-CRNN        & 11.061             & 0.441          \\
\hline
\end{tabular}
\label{tab:PFD}
\vspace{-12pt}
\end{table}

\subsection{Multi-Frequency Dynamic Convolution}
To test the effect of multiple FDY conv modules, we experimented with MDFD conv without dilation in this subsection. We call this module as multi-frequency dynamic convolution (MFD conv). The result is shown in Table \ref{tab:MFD_one-eighth}, where \#DYbr stantds for the number of dynamic branches. Four dynamic branches with proportion of $1/32$ and two dynamic branches with proportion of $1/16$-sized are experiemnted to test if multiple dynamic branches which sum up to $1/8$ of output channels are beneficial, but the results show that they cause performance drop thus dynamic branch with proportion of $1/8$ is minimum that does not harm the performance compared to FDY conv. Furthermore, we experimented on multiple $1/8$-sized FDY modules for further performance enhancement.The results state that five and six branches improve SED performance the most, while they are still lighter than FDY conv model. This results demonstrates that multiple dynamic branches enhances SED performance, possibly by learning different dynamic pattern recognition by each branch. Also, we infer from the results that static branch channel proportion is $1/4\sim3/8$ are minimally required since less proportion of static branch deteriorates the performance.
\begin{table}[t]
\caption{Performance of multi-frequency dynamic convolutions models.}
\centering
\begin{tabular}{c|cc|c}
\hline
\textbf{models} & \textbf{\# DYbr} & \textbf{Params(M)} & \textbf{PSDS1}  \\ 
\hline
FDY-CRNN        & 1                & 11.061             & 0.441        \\
PFD-CRNN (1/8)  & 1                & 5.401              & 0.442         \\ 
\hline
MFD-CRNN (1/32) & 4                & 5.896              & 0.430         \\
MFD-CRNN (1/16) & 2                & 5.566              & 0.439         \\
MFD-CRNN (1/8)  & 2                & 6.374              & 0.439         \\ 
MFD-CRNN (1/8)  & 3                & 7.348              & 0.444          \\ 
MFD-CRNN (1/8)  & 4                & 8.322              & 0.440         \\
MFD-CRNN (1/8)  & 5                & 9.296              & \textbf{0.449} \\ 
MFD-CRNN (1/8)  & 6                & 10.270             & \textbf{0.452} \\ 
MFD-CRNN (1/8)  & 7                & 11.243             & 0.445         \\
MFD-CRNN (1/8)  & 8                & 12.217             & 0.447          \\ 
\hline
\end{tabular}
\label{tab:MFD_one-eighth}
\vspace{-12pt}
\end{table}

\begin{table}[t]
\caption{Performance of multi-dilated frequency dynamic convolution models with varying dilation size sets. All partial dynamic branches have proportion of 1/8.}
\centering
\begin{tabular}{c|c|c}
\hline
\textbf{models} & \textbf{Dilation Sizes}             & \textbf{PSDS1} \\ 
\hline
FDY-CRNN  & (1)                               & 0.441        \\
PFD-CRNN  & (1)                               & 0.442          \\ 
MFD-CRNN  & (1)$\times$5                      & \textbf{0.449} \\ 
\hline
MDFD-CRNN &(1)$\times$4+(2)                   & 0.449          \\
MDFD-CRNN &(1)$\times$4+(3)                   & 0.448          \\
MDFD-CRNN &(1)$\times$4+(2,2)                 & 0.448          \\
MDFD-CRNN &(1)$\times$4+(2,3)                 & \textbf{0.451} \\
MDFD-CRNN &(1)$\times$4+(3,3)                 & 0.446          \\
MDFD-CRNN &(1)$\times$4+(2,2,3)               & 0.446          \\
MDFD-CRNN &(1)$\times$4+(2,3,3)               & \textbf{0.451} \\
\hline
MDFD-CRNN &(1)$\times$3+(2,3)$\times$2        & 0.448          \\
MDFD-CRNN &(1)$\times$3+(2,2,3)$\times$2      & 0.449          \\ 
MDFD-CRNN &(1)$\times$3+(2,3,3)$\times$2      & 0.450          \\
MDFD-CRNN &(1)$\times$3+(2,3)+(2,3,3)         & \textbf{0.451} \\
\hline
MDFD-CRNN &(1)$\times$2+(2,3)$\times$3        & 0.449          \\
MDFD-CRNN &(1)$\times$2+(2,2,3)$\times$3      & \textbf{0.452} \\
MDFD-CRNN &(1)$\times$2+(2,3,3)$\times$3      & 0.447          \\
MDFD-CRNN &(1)$\times$2+(2,3)+(2,2,3)+(2,3,3) & 0.447          \\ 
\hline
\end{tabular}
\label{tab:MDFD_5}
\vspace{-12pt}
\end{table}
 \begin{table*}[ht]
\caption{Performance of multi-dilated frequency dynamic convolution models with pre-convolution and varying channel sizes.}
\centering
\begin{tabular}{c|ccc|c}
\hline
\textbf{models} & \textbf{\# Channels} & \textbf{Dilation Sizes}                                 & \textbf{Params(M)} & \textbf{PSDS1} \\
\hline
FDY-CRNN        & 8/8 (32,64,128,256)  & (1)                                                     & 11.061             & 0.441          \\
DFD-CRNN        & 8/8 (32,64,128,256)  & (2,3)                                                   & 11.061             & 0.448          \\
PFD-CRNN        & 8/8 (32,64,128,256)  & (1)                                                     & 5.401              & 0.442          \\
MFD-CRNN        & 8/8 (32,64,128,256)  & (1)$\times$5                                            & 9.296              & 0.449          \\ 
MDFD-CRNN      & 8/8 (32,64,128,256)  & (1)$\times$2+(2,3)+(2,2,3)+(2,3,3)                      & 9.296              & 0.447          \\ 
\hline
FDY-CRNN        & 11/8 (44,88,176,352) & (1)                                                     & 19.317             & 0.434          \\ 
MDFD-CRNN       & 11/8 (44,88,176,352) & (1)$\times$8                                            & 18.157             & 0.449          \\ 
MDFD-CRNN       & 11/8 (44,88,176,352) & (1)$\times$3+(2)+(3)+(2,3)+(2,2,3)+(2,3,3)              & 18.157             & 0.454          \\
MDFD-CRNN       & 11/8 (44,88,176,352) & (1)$\times$5+(2,3)+(2,2,3)+(2,3,3)                      & 18.157             & \textbf{0.455} \\
MDFD-CRNN       & 11/8 (44,88,176,352) & (1)$\times$6+(2,3)+(2,2,3)+(2,3,3)                      & 19.582             & 0.450          \\
MDFD-CRNN       & 13/8 (52,104,208,416)& (1)$\times$5+(2,2)+(3,3)+(2,3)+(2,2,3)+(2,3,3)          & 26.191             & 0.446          \\ 
MDFD-CRNN       & 14/8 (56,112,224,448)& (1)$\times$5+(2,2)+(3,3)+(2,2,3,3)+(2,3)+(2,2,3)+(2,3,3)& 30.894             & 0.433          \\ 
\hline
\end{tabular}
\label{tab:preMDFD}
\vspace{-12pt}
\end{table*}
\subsection{Multi-Dilated Frequency Dynamic Convolution}
By introducing dilation to the MFD-CRNN model with five dynamic branches which showed good size to performance balance, we experimented on dilating basis kernels in some of five dynamic branches in order to diversify the roles by different dynamic branches \cite{DFD}. The results are shown in Table \ref{tab:MDFD_5}, where dilation sizes are notated to show only frequency-wise dilation. One parenthesis set describes dilation size set of single dynamic branch where (1) refers to non-dilated dynamic branch and otherwise refes to dilation sizes of basis kernels those are dilated. For example, $(1)\times3+(2,3)\times2$ means 3 non-dilated dynamic branches and 2 dynamic branches have two non-dilated basis kernels, one basis kernel with dilation size of 2 and the last basis kernel with dilation size 3. Note that all dynamic branches consist of four basis kernels. Dilating basis kernels of dynamic branches result in similar or slightly better performance by \cite{DFD}. We interpret the results that as much as DFD conv branch benefits SED, FDY conv branch also contributes well.

\subsection{Multi-Dilated Frequency Dynamic Convolution with Pre-convolution and Varying Channel sizes}
We experimented the effect of additionally introducing dynamic branches with expanded channel sizes in Table \ref{tab:preMDFD}. In addition, to further enhance the performance, pre-convolution module with output channel size of 16 is added in front of CNN module to add MDFD conv to 1st convolution layer. Without pre-convolution, the input channel size of 1st convolution layer is one, thus it is meaningless to extract $K$ frequency-adaptive attention weights from single channel. Addition of three dilated dynamic branches with varying dilation sizes results in PSDS1 of 0.455, which outperforms FDY-CRNN by 3.17\%. However, introducing more dilated dynamic branches results in worse performance, showing that too many dynamic branches rather harms the performance. While it involves far more parameters, the results show that several dynamic branches without dilation helps the performance and adding right amount of dilated dynamic branches enhances the performance.

\begin{table}[t]
\caption{Performance comparison of MDFD-CRNN with state-of-the-art models wihtout external dataset. Pp, mf cw and ssl stand for post processing, median filter, class-wise and semi-supervised learning method choice.}
\centering
\begin{tabular}{c|cc|cc}
\hline
\textbf{models}       & \textbf{PP}  & \textbf{SSL} & \textbf{PSDS1} & \textbf{tPSDS1}\\ 
\hline
FDY-CRNN \cite{FDY}   & cw-mf        & MT           & 0.451          & -              \\
DFD-CRNN \cite{DFD}   & cw-mf        & MT           & 0.455          & 0.465          \\
MFD-CRNN \cite{mdfdy} & cw-mf        & MT           & 0.461          & -              \\
MFD-CRNN \cite{mdfdy} & cw-mf        & CMT          & \textbf{0.470} & -              \\
\hline
MDFD-CRNN             & mf           & MT           & 0.455          & 0.468          \\ 
MDFD-CRNN             & cw-mf        & MT           & 0.461          & 0.474          \\ 
MDFD-CRNN             & cSEBBs       & MT           & -              & \textbf{0.485} \\ 
\hline
\end{tabular}
\label{tab:sotawdext}
\vspace{-12pt}
\end{table}

\subsection{Comparison with State-of-the-art Models}
We compared best MFD-CRNN model with state-of-the-art model without external dataset on Table \ref{tab:sotawdext}. With class-wise median filter, it achieves same score with MFD-CRNN \cite{mdfdy}. Combining MFD cov and confident mean teacher methods to MDFD-CRNN could further enhance the performance, but it is out of scope of this work. Application of cSEBBs on MDFD-CRNN achieves true PSDS1 of 0.485, which is the state-of-the-art score on DESED without external dataset \cite{SEBB, truePSDS}. True PSDS refers to threshold-independent PSDS by \cite{truePSDS} in this work, opposed to the PSDS that uses 50 thresholds as in DCASE 2021 and 2022 baseline \cite{DCASEtask4}.

To compare the performance of MDFD conv with state-of-the-art models with pre-trained audio models, we implemented MDFD conv on DCASE 2024 challenge Task 4 setting as shown in Table \ref{tab:dcase} \cite{atstsed, audioset, beats, ATST}. Table \ref{tab:dcase} includes other state-of-the-art models with pre-trained models but without ensembling and self-training with AudioSet, since those methods are mainly for maximizing the performance rather than for comparing effect of methods. Detailed settings of ATST-frame, BEATs and CRNN (ABC) model are referred to \cite{dcase2024mytechrep}. Change-detection-based sound event bounding boxes (cSEBBs) is used as post-processing \cite{SEBB}. Evaluation metric used is true PSDS1 \cite{truePSDS}. Fine-tuning of ATST-frame further enhanced the performance, while it does not show significant performance gain with cSEBBs. This could be due to use of class-wise median filter. The resultant model outperforms DCASE 2023 and 2024 challenge winners without ensemble and self training \cite{dcase2023a_1st, dcase2024_1st}. Also, ABC w/ MDFD-CRNN best model performs close to ATST-SED model with fine-tuning ATST-frame \cite{atstsed}.

\begin{table}[t]
\caption{Performance comparison of PFD, MDFD-CRNN with state-of-the-art models with pretrained audio models without AudioSet or ensemble. Pp, mf cw, and ft stand for post processing, median filter, class-wise and fine-tuning. ABC stands for model with ATST-frame, BEATs and CRNN.}
\centering
\begin{tabular}{c|c|c}
\hline
\textbf{models}                                 & \textbf{PP} & \textbf{tPSDS1}\\ 
\hline
FDY-LKA-CRNN + BEATs  \cite{dcase2023a_1st}     & cw mf       & 0.525          \\
CRNN + ATSTframe \cite{atstsed}                 & cw mf       & 0.492          \\
CRNN + ATSTframe + ft \cite{atstsed}            & cw mf       & \textbf{0.583} \\
CRNN + BEATs + ft \cite{dcase2024_1st}          & cw mf       & 0.539          \\
\hline
ABC                                             & mf          & 0.507          \\ 
ABC w/ PFD-CRNN                                 & mf          & 0.517          \\
ABC w/ MDFD-CRNN.                               & mf          & 0.524          \\ 
ABC w/ MDFD-CRNN + ft                           & cw mf       & \textbf{0.550} \\ 
\hline
ABC                                             & cSEBBs      & 0.546          \\ 
ABC w/ PFD-CRNN                                 & cSEBBs      & 0.558          \\
ABC w/ MDFD-CRNN                                & cSEBBs      & \textbf{0.577} \\ 
ABC w/ MDFD-CRNN + ft                           & cSEBBs      & 0.554          \\ 
\hline
\end{tabular}
\label{tab:dcase}
\vspace{-12pt}
\end{table}

\section{Conclusion}
\label{sec:conclusion}
Proposed partial frequency dynamic convolution concatenates outputs of FDY conv module and convolution module to reduce the model size. Consequently, PFD-CRNN reduced the number of parameters by 51.9\% compared to FDY-CRNN while retaining the performance. Proposed multi-frequency dynamic convolution module introduces more dynamic branches to further enhance the performance. Best MDFD-CRNN model outperforms FDY-CRNN by 3.17\%. Extensive ablation studies showed that minimum proportion of dynamic branches is $1/8$, minimum proportion of static branch is $1/4\sim3/8$, and non-dilated FDY conv branches also beneficial thus additional DFD conv branches with expanded channels further enhanced the performance. With cSEBBs, MDFD-CRNN achieves true PSDS1 of 0.485, which is state-of-the-art in DESED without external dataset. MDFD-CRNN with pre-trained models shows comparable performances to previous state-of-the-art models.

\bibliographystyle{IEEEtran}
\bibliography{refs}

\end{document}